\documentclass[superscriptaddress, prl,a4paper,12pt]{revtex4}
 
\pdfoutput=1 
\usepackage{float}
\usepackage{fancyhdr}
\usepackage{color}
\usepackage{graphicx}
\usepackage{epsfig} 
\usepackage{amssymb}
\usepackage{amsmath,amsfonts,verbatim}
\usepackage{booktabs}
\usepackage{hyperref}
\usepackage{eufrak}
\usepackage{mathrsfs}
\usepackage[footnotesize,singlelinecheck=off,justification=raggedright]{caption}
\def\Bra#1{\left<1>}

\DeclareMathAlphabet{\mathbbmsl}{U}{bbm}{m}{sl}
{\catcode`\|=\active\gdef\Braket#1{\left<\mathcode`\|"8000\let|\bravert {#1}\right>}}
\def\bravert{\egroup\,\vrule\,\bgroup}

\begin{document}

\title{Unraveling the Landau's consistence criterion and the meaning of interpenetration in the ``Two-Fluid'' Model.}
\author{J. L. Alonso}
\affiliation{Departamento de F\'{\i}sica Te\'orica, Universidad de Zaragoza,
50009 Zaragoza, Spain}
\affiliation{Instituto de Biocomputaci\'on y F\'{\i}sica de Sistemas Complejos (BIFI), Universidad de Zaragoza,
50018 Zaragoza, Spain}
\affiliation{Unidad asociada IQFR-BIFI, 50018 Zaragoza, Spain}
\author{F. Ares}
\email{ares@unizar.es}
\affiliation{Departamento de F\'{\i}sica Te\'orica, Universidad de Zaragoza,
50009 Zaragoza, Spain}

\author{J. L. Brun}
\affiliation{Departamento de F\'{\i}sica Te\'orica, Universidad de Zaragoza,
50009 Zaragoza, Spain}

\begin{abstract}
 In this letter we show that it is possible to unravel both the 
 physical origin of the Landau's consistence criterion 
 and the specific and subtle meaning of interpenetration of the
 ``two fluids'' if one takes into account 
 that in the hydrodynamic regime one needs a 
 coarse-graining in time to bring the system into local
 equilibrium. That is, the fuzziness in time is relevant
 for the phenomenological Landau's consistency criterion and the meaning 
 of interpenetration. Note also that we are not questioning the validity
 of the ``Two-Fluid'' Model. 
\end{abstract}

\maketitle
\section{I. Introduction}

A quantum fluid is a substance that remains fluid (i.e. gas or liquid) 
at such low temperatures that the effects of quantum mechanics play a dominant 
role and the laws of classical statistical mechanics do not serve.  
The two isotopes of Helium, $^4$He and $^3$He, are the only quantum 
fluids on the Earth. The $^4$He atom is a zero-spin boson, which is 
determinant for its quantum properties. In fact, at temperatures 
and pressures below $T_\lambda\approx 2.17$ K and 2.5 MPa respectively, 
$^4$He is at the so called He II superfluid phase where it displays 
among other properties \cite{Nozieres, Tilley} that of flowing through 
narrow channels with no measurable viscosity.

In order to explain the behaviour of He II, L. Tisza and L.D. Landau 
\cite{Tisza, Landau, Landau2} developed the Two-Fluid Model. 
They described the He II phase by two components, one viscous and 
another non-viscous, related to each other in a very sophisticated way, 
as we shall explain in Section II. 
The main differences between the Tisza and Landau proposals are discussed 
in detail in \cite{Balibar}. Here we will deal with the theory of Landau 
and Feynman \cite{Landau,Landau2,Feynman1,Feynman2,Feynman3}, 
the commonly accepted Two-Fluid Model. 

A very complete summary of the Two-Fluid theory can be seen in \cite{Donnelly}.
In \cite{Amigo} the authors grasp some interesting subtleties of the  Two-Fluid model.
We only highlight here that one of the most important Feynman's contributions 
was to note \cite{Feynman1,Feynman2} the relevance of Bose-Einstein
statistics to correctly account for the thermally excited states of compression, 
with a resultant change in the density: the phonon-roton density 
modes of the normal component (see also \cite{Bossy}).

\section{II. The Landau's consistency criterion and the meaning of interpenetration}

In the Two-Fluid Model of the He II, the thermodynamical equilibrium 
states are described, at any temperature $T < T_\lambda$, \cite{Leggett}, 
by two independent velocities, $\vec{v}_s(\vec{r})$, $\vec{v}_n(\vec{r})$, 
and associated densities, $\rho_s(\vec{r})$ and $\rho_n (\vec{r})$, 
such that the densities of mass current $\vec{J}(\vec{r})$ and of kinetic 
energy flow $Q(\vec{r})$ are given by,
\begin{equation}\label{eq:1}
\rho(\vec{r}) = \rho_ s (\vec{r} ) + \rho_n (\vec{r}),                      
\end{equation}                                 
\begin{equation}\label{eq:2}
J(\vec{r})= \rho_s (\vec{r})\vec{v}_s (\vec{r})
+\rho_n(\vec{r})\vec{v}_n (\vec{r}),
\end{equation}
\begin{equation}\label{eq:3} 
Q(\vec{r}) = \frac{1}{2}\left[\rho_s(\vec{r})\vec{v}_s(\vec{r})^2 
+ \rho_n (\vec{r})v_n^2 (\vec{r})\right],
\end{equation}
where $\rho(\vec{r})$ is the total mass density and the 
superfluid component $\rho{ρ}_s(\vec{r})$ is not viscous. 
The entropy of the liquid is entirely attributed to the normal 
fluid part $\rho_n$, 
$$\rho(\vec{r}) S= \rho_n(\vec{r}) S_n,$$ 
where $S$ is the total entropy per mass of the liquid and 
$S_n$ the normal fluid entropy. 

At $T=0$, the entire liquid is supposed to be superfluid, 
i.e. $\rho=\rho_s$ while at $T= T_\lambda$ 
the superfluid component vanishes, i.e. $\rho=\rho_n$.

The model is completed by taking into account the following Landau's 
consistency criterion \cite{Landau, Landau2}: ``...there is no division 
of the real particles of the liquid into ``superfluid” and ``normal" ones. 
In a certain sense one can speak of ``superfluid" and ``normal" masses 
of liquid as of masses connected with two simultaneously possible movements, 
but this by no means signifies the possibility of a real division of the 
liquid into two parts".

Here we shall discuss on the Landau's consistency criterion
and on the meaning of interpenetration.

Firstly, $\vec{r}$ must be understood, as always in fluid mechanics 
\cite{Landau3}, ``as the position of the volume element corresponding 
to a ``fluid particle", in the sense of a volume element containing 
many particles though regarded as a point". 
Then, when we speak of the displacement of a fluid particle we mean 
not the displacement of an individual atom of He, but that of a volume 
element containing many atoms, though still regarded as a point. 
Since both densities (\ref{eq:1}) are referred to the same 
$\vec{r}$ , they ``lend nicely to an intuitive picture of He II as a 
mixture of two independent, interpenetrating ``fluids" or ``components", 
the ``superfluid" and the ``normal" components" \cite{Leggett}. 
Of course, the interpenetration is not just that both components refer to 
the volume element of the fluid positioned in $\vec{r}$ , this is common 
to fluid dynamics for a standard mixture of fluids, as e.g. water and wine, 
(see Chapter VI of \cite{Landau3}).
In order to unravel the physical meaning of the interpenetration
one can take into account a coarse-graining in time in correspondence
with the mentioned volume element, or coarse-graining in space, positioned 
in $\vec{r}$. 

Indeed, following the fundamental basis of the hydrodynamic regime 
\cite{Mota, Koide, Derradi, Denicol, Pu}, one should introduce a 
coarse-graining in time. In fact, 
Landau and Lifshitz thought that the fuzziness in time was somewhat irrelevant. 
Then, assuming this irrelevance, they extended the 
Navier-Stokes equation to a relativistic covariant form (see Chapter XV 
of \cite{Landau3}). 
However, this equation is known to be unstable, and we now know its origin: 
any volume element requires a finite relaxation time, or coarse-graining 
in time, to reach a thermodynamical equilibrium and the Landau and Lifshitz 
naive covariant extension of the Navier-Stokes equation, obtained neglecting
this necessary finite relaxation time, contains acausal modes which are 
the origin of the instability. Let us recall Section 26 of \cite{Landau3}: 
``Not every solution of the equations of motion, even if it is exact, 
can actually occur in Nature. The flows that occur in Nature must 
not only obey the equations of fluid dynamics, but also be stable". 
That is, the consistency of the theory advises us to consider a coarse-graining
in time. Indeed, in the hydrodynamic regime one needs a coarse-graining 
in time larger than the relaxation time that brings the system into local 
equilibrium.

It is true that for the most of non-relativistic cases we do not need 
explicitly a coarse-graining in time, as in \cite{Landau3}. Nevertheless, now 
we will see that taking into account this coarse-graining 
we can unravel the physical origin of the Landau's consistency criterion
and the meaning of interpenetration.

In fact, when fluid mechanics describes the evolution of a fluid with time 
what is being described is what it would be observed in successive snapshots 
corresponding to successive coarse-graining times. In our case what it will 
be observed is that the He atoms participating in the phonon-roton density 
modes \cite{Feynman1,Feynman2,Bossy} of the ``normal component" will be 
different at each given coarse-graining time (as illustration one can think in
the thermodynamic description of vapor-liquid equilibrium: molecules that 
participate in the vapor and in the liquid change with time). 
Then, ``...there is no division of the real particles of the liquid into 
``superfluid" and ``normal" ones..." or ``...there is no chance of a real 
division of the liquid into two parts", as Landau emphasizes. 
This is a crucial difference with the water/wine mixture case mentioned 
above, in which there are water molecules different from wine molecules.
To sum up, in the two fluid model the He atoms can not be labeled according 
to their contribution to the superfluid, $\rho_s(\vec{r})$, or to the normal 
component, $\rho_n(\vec{r})$: they are interpenetrating fluids, as 
Leggett says.

\section{III. The local equilibrium hydrodynamic regime (LEHR)}

The most extraordinary fact of the superfluid phase of He II is the
stability of its flow. Since the discovery of this dramatic feature,
it was described (almost always) within the framework of the LEHR, at
the beginning implementing in this frame the Tisza's two fluid model
and later the Landau's model. In section II we have taken advantage of
this fact, in particular of the physical finite relaxation time,
fundamental ingredient of the LEHR. Superflow stability has also been
studied by starting from the hypothesis that steady supercurrents are
metastable states. Therefore, in this case, it must be discussed in
the language of statistical mechanics of irreversible processes
\cite{Tilley, Langer}.

Now, on the one hand, we have the physical stability of the superflow
and the physically meaningful relaxation time already discussed in
section II. On the other hand we can ask about the absence of
numerical instabilities when one is working with a system of nonlinear
time-dependent equations, where also a mathematical $\Delta t$, without
physical meaning, appears. The comment on this is the reason for this
section.

Let us recall that in the two-fluid models of helium II, while at low
enough flow velocities the normal and superfluid components of helium
II move independently, at higher velocities ( see section "Two-Fluid
Models" of the second article of \cite{Donnelly}), "quantized vortex 
lines appear in the superfluid component and the two fluids become 
coupled by a force called mutual friction". Phonon-rotons density modes 
continue being the normal component. To be specific, let's consider the
Landau's Two Fluid Model (LTFM) in this framework.

Two points to keep in mind are:

First, the LEHR is also the physical frame suitable for raising
generalizations of the LTFM (see the first three sections of \cite{Geurst}).

Second, the concrete physical properties which characterize a system
in the LEHR must be searched using physical arguments as, for example,
in the case considered, what is the physical time scale for the vortex
lines to equilibrate? A vortex line is a hole with quantised
circulation, so the answer must have to do with the speed of sound
(240 m/s in helium), the size of the hole (diameter about 10$^{-10}$ m)
or the distance between vortices. Although much is known about
quantized vortices in Helium II \cite{Donnelly2}, as far as we know, that
knowledge is not sufficient to answer this question. This lack of
knowledge or certainty does not prevent us the construction of models
living thanks to the physical support given by the LEHR: these models
are supported on their success \cite{Landau3, Landau, Landau2, Feynman3, Feynman1, Feynman2}.

We conclude this section by summarizing the situation from the
mathematical side. To prove mathematically the absence of numerical
instabilities when one is working with a system of nonlinear
time-dependent partial differential equations in two or three spatial
dimensions, as the LTFM equations, is a very difficult problem. In
fact, as can be seen in \cite{Henderson, Barenghi, Barenghi2}, in 
order to achieve numerical stability, one monitors the power spectrum of the
solution to make sure that it decays at large spatial frequency.The
power-law decay means that spectral convergence has been achieved.
Here the only point we want to clarify is that, in these studies, the
time step $\Delta t$, which sets the boundary between
stable and unstable numerical solutions (see section 5 of \cite{Henderson}), 
means nothing physically, because it depends on the precise
numerical method: this $\Delta t$ is not a physically meaningful
relaxation time (see Section II).

\section{IV. Conclusion}

The discussion in Section II is supported on the physical finite
relaxation time required by the fundamental basis of the LEHR, in
which the LTFM and its generalization live. As we have seen,
neglecting it can lead to serious inconsistencies and sometimes to
forget the possible physical origin of important ingredients of a
phenomenological theory such as, in our case, the Landau's consistency criterion and
the meaning of interpenetration in the Two-fluid model.

Finally, note that one can conclude that the name of the model is in fact a bit
misleading and it should contain a small footprint of the Landau's consistency 
criterion by typing in quotes ``two fluid", as we have done in the title. 
\newline
\newline
\textit{Acknowledgments:} We gratefully thank Prof. C.F. Barenghi, Prof.
T.Kodama and Prof. A.J. Leggett for correspondence exchange and Prof.
A. Cruz for very illuminating discussions. In addition, Prof. Kodama
has helped us to complete our knowledge about the hydrodynamic regime
and without the help of Prof. Barenghi we would not have got all the
content of his works, essential for the discussion in Section III.
We also acknowledge the referee's recommendations. 
This work was supported by Diputaci\'on General de Arag\'on (Spain)-FSE
Grants E24/2 and E24/3 and by Ministerio de Econom\'{\i}a y Competitividad 
(Spain) Grants FIS2017-82426-P and FPA2015-65745-P. 
FA is supported by the DGA/European Social Fund Grant C070/2014.
\newline
\newline
\textit{Author contribution:} all the authors have contributed equally to the 
work.


\begin{thebibliography}{XXX}
    
    \bibitem{Nozieres} P. Nozi\`eres and D. Pines, ``The Theory of Quantum Liquids", 
    1999, Perseus Books, Cambridge, Massachusetts.

    \bibitem{Tilley} D.R. Tilley and J. Tilley, ``Superfluidity and Superconductivity" 
    (Third Edition,1990), Adam Hilger, Bristol and New York.
    
    \bibitem{Tisza} L. Tisza, ``On the thermal supraconductibility of liquid helium II 
    and the Bose-Einstein statistics", C.R. Acad. Sci. 207 (1938) 1035
    
    \bibitem{Landau} L.D. Landau, ``The theory of superfluidity of Helium II",
    J. Phys. U.S.S.R. 5, 71 (1941).

    \bibitem{Landau2} L.D. Landau, ``On the theory of superfluidity of Helium 
    II", J. Phys. U.S.S.R. 11,91 (1947).
    
    \bibitem{Balibar} S. Balibar, ``The Discovery of Superfluidity", 
    Journal of Low Temperature Physics,146 (2007) 441, S. Balibar 
    ``Laszlo Tisza and the two-fluid model of superfluidity", Comptes Rendus Physique 18 (2017) 586-591.
    
    \bibitem{Feynman1} R.P. Feynman, ``Application of Quantum Mechanics to Liquid Helium" 
    in ``Progress in low Temperature Physics", Vol.I, North-Holland, Amsterdam, (1955).
    
    \bibitem{Feynman2} R.P. Feynman,  ``Statistical Mechanics: A set of Lectures", Benjamin, Readin, MA, (1982).
    
    \bibitem{Feynman3} R.P. Feynman, ``Atomic Theory of the 2-fluid model of liquid Helium" Phys. Rev. 94(1954)262
    
    \bibitem{Donnelly} R.J. Donnelly, ``The two-fluid theory and second sound 
    in liquid helium", Phys. Today 62, 34 (2009), ``Rotons: a low-temperature puzzle", Phys. World (1997) 10 (2) 25, 
    and ``The Discovery of Superfluidity", Phys. Today 48, 30 (1995).
    
    \bibitem{Amigo} M. L. Amig\'o, T. Herrera, L. Ne\~ner, L. Peralta Gavensky, 
    F. Turco, and J. Luzurriaga, ``A quantitative experiment on the fountain effectiveness in superfluid
    Helium'', Eur. J. Phys. 38 (2017) 055103.

    \bibitem{Bossy} J. Bossy, J. Ollivier, H. Schober and H.R. Glyde, 
    ``Phonon-Roton modes in liquid He coincide with Bose Einstein condensation" EPL, 98 (2012) 56008.
    
    \bibitem{Leggett} A.J. Leggett, ``Quantum liquids: Bose condensation and 
    Cooper pairing in condensed matter systems", Oxford University Press, 2006.

    \bibitem{Landau3} L.D. Landau and E.M. Lifshitz, ``Fluid Mechanics", 
    Pergamon Press (1959).

    \bibitem{Mota} Ph. Mota, T. Kodama, R. Derradi de Souza and J. Takahashi, 
    ``Coarse-graining scale and effectiveness of hydrodynamic modeling", 
    Eur. Phys. J. A (2012) 48:165.

    \bibitem{Koide} T. Koide, G. S. Denicol, Ph. Mota, and T. Kodama,
    ``Relativistic dissipative hydrodynamics: A minimal causal theory", 
    Phys. Rev. C 75 (2007) 034909.

    \bibitem{Derradi} R. Derradi de Souza, T. Koide and T. Kodama,
    ``Hydrodynamic approaches in relativistic heavy ion reactions", 
    Progress in Particle and Nuclear Physics 86 (2016) 35-85.

    \bibitem{Denicol} G. S. Denicol, T. Kodama, T. Koide and Ph. Mota, 
    ``Stability and causality in relativistic dissipative hydrodynamics",
    J. Phys. G: Nucl. Part. Phys. 35 (2008) 115102.

    \bibitem{Pu} S. Pu, T. Koide, and D. H. Rischke, 
    ``Does stability of relativistic dissipative fluid dynamics imply 
    causality?", Phys. Rev. D 81, 114039 (2010).
    
    \bibitem{Langer} J. S. Langer and J.D. Reppy, 
    "Intrinsic critical velocities in superfluid helium II", 
    Progress in Low Temperature Physics Volume 6, 1970, pag. 1 (1970).
    
    \bibitem{Geurst} J.A.Geurst,  "General theory unifying and
     extending the Landau-Khalatnikov, Ginzburg-Pitaevskii, and
     Hills-Roberts theories of superfluid $^4$He'',
     PRB, 22, 3207 (1980). 

    \bibitem{Donnelly2} R.J. Donnelly, 
    "Quantized vortices in helium II", Cambridge
     University Press" (1991).

    \bibitem{Henderson}  Karen L. Henderson, Carlo F. Barenghi, 
    "Numerical Methods for Two-Fluid Hydrodynamics: Application 
    to the Taylor Vortex Flow of Superfluid Helium II", 
    Journal of Low Temperature Physics 98, 351 (1995).

    \bibitem{Barenghi} Carlo F. Barenghi,  C. A. Jones, 
    "The stability of the Couette flow of helium II", 
    J . Fluid Mech. (1988), vol. 197, p p . 551-569.

    \bibitem{Barenghi2} Carlo F. Barenghi, 
    "Vortices and the Couette Sow of helium II", 
    PRB 45, 2290.

  

\end{thebibliography}
\end{document}